\documentclass[twocolumn,aps,superscriptaddress,longbibliography,amsmath,amssymb]{revtex4-1}
\usepackage{graphicx}
\usepackage{dcolumn}
\usepackage{simplewick}
\usepackage{siunitx}
\usepackage{bm}
\usepackage{times}
\usepackage{hyperref}
\usepackage[usenames,dvipsnames]{color}
\hypersetup{
  colorlinks=false,
  colorlinks=true,
  citecolor=blue,
  linkcolor=blue,
  urlcolor=blue}

\begin{document}

\title{Fock-space relativistic coupled-cluster calculation of hyperfine induced 
       $\bf {^1S_0 \rightarrow {^3P^o_0}}$ clock transition in Al$^+$ }

\author{Ravi Kumar}
\affiliation{Department of Physics, Indian Institute of Technology,
             Hauz Khas, New Delhi 110016, India}

\author{S. Chattopadhyay}
\affiliation{Department of Physics, Kansas State University,
             Manhattan, Kansas 66506, USA
             }

\author{D. Angom}
\affiliation{Physical Research Laboratory,
             Ahmedabad - 380009, Gujarat,
             India}

\author{B. K. Mani}
\affiliation{Department of Physics, Indian Institute of Technology,
              Hauz Khas, New Delhi 110016, India} 

\begin{abstract}

  We have developed an all-particle Fock-space relativistic coupled-cluster 
method to calculate the properties of two-valence atoms and ions. Using the 
method we compute the properties associated with hyperfine induced 
$^1S_0 - ^3P^o_0$ clock transition in Al$^+$. Our result of the $^3P^o_0$ 
metastable state life time, $20.20 \pm 0.91$ s, is in excellent agreement 
with the experimental value, $20.60 \pm 1.4$ s [Phys. Rev. Lett. {\bf 98}, 220801 (2007)]. 
Our studies show that the contributions from the triple excitations, and the corrections 
from the Breit interaction and QED effects are essential to obtain accurate 
clock properties in Al$^+$.
\end{abstract}

\maketitle
\section{Introduction}

  Development of atomic clocks as frequency standard provide a roadmap to study 
fundamental as well as technological applications. Some important examples are 
the variation of the fundamental constants, probing physics beyond the standard 
model of particle physics, navigation systems and the basis for the 
redefinition of the second 
\cite{karshenboim-10,grewal-13,safronova-18,riehle-18}. 
The recent frequency standard 
experiments \cite{rosenband-07,chou-10,chen-17,brewer_19a,brewer_19b} 
in optical domain have reported $^1S_0-^3\!P^o_0$ transition in Al$^+$ 
as one of the most accurate clock transitions. Though the $^1S_0-^3\!P^o_0$ 
transition is highly forbidden based on the selection rule of the total 
electronic angular momentum $J$, it is possible through hyperfine mixing 
of the $^3P^o_0$ state with $^3P^o_1$ and $^1P^o_1$ states. The life time of 
the $^3P^o_0$ metastable clock state was measured with high accuracy by 
Rosenband and collaborators \cite{rosenband-07} using the quantum logic 
spectroscopy technique. Three key factors favoring the choice for this 
transition as clock transition are low sensitivity to electromagnetic fields, 
narrow natural linewidth and small room temperature black-body radiation 
shift. The latter is due to small difference between the polarizabilities of 
$^1S_0$ and $^3P^o_0$ states \cite{kallay-11,safronova-11}. A recent work 
reported the fractional frequency uncertainty of a $^1S_0-^3\!P^o_0$ transition 
based Al$^+$ clock as $9.4\times10^{-19}$ \cite{brewer_19a}. 
And, this, perhaps, is the most precise atomic clock in existence today.

  Despite the important applications of the $^1S_0-^3\!P^o_0$ 
hyperfine induced electric dipole transition ($E1_{\rm HFS}$), and several 
experimental investigations in progress, very little theoretical data on the 
associated properties is available. For example, there are only two results on 
the life time of the $^3P^o_0$ metastable clock state \cite{brage-98,kang-09}, 
and both are based on the method of
multiconfiguration Dirac-Fock (MCDF). To the best of our knowledge, there are 
no theoretical results using the accurate many-body methods like relativistic 
coupled-cluster (RCC). It is to be emphasized that the RCC is considered 
to be one of the most accurate many-body theories for the structure and 
properties calculations of atoms and ions. It accounts for the electron 
correlation effects to all-orders of residual Coulomb interaction, and has 
been employed to obtain accurate properties in several closed-shell 
and one-valence atoms and ions \cite{pal-07,mani-09,nataraj-11,ravi-20}. 
The implementation of RCC for two-valence atomic systems is, however, limited 
to few studies \cite{eliav-95,eliav-95b,mani-11}. The reason, perhaps, is the 
complications associated with its implementation for two-valence systems. 
To be more precise, there are three main hurdles. First, due to the 
multireference nature of the configuration space, the model wave function is 
not well defined. This needs a special treatment through the 
diagonalization of the effective Hamiltonian matrix. Second, the atomic states 
are the eigen states of the total angular momentum, which leads to a 
complication in the angular factors associated with antisymmetrized 
many-electron states. And third, divergence due to {\em intruder} states. 

  It can thus be surmised that there is a clear research gap in terms of
the scarcity of accurate theoretical data on the $^1S_0-^3P^o_0$ transition 
properties. The aim of this work is to fill this research gap. 
To address this in a comprehensive way, we adopt a three 
prong approach. First, we develop a Fock-space relativistic coupled-cluster 
(FSRCC) based method for structure and properties calculations of two-valence
atoms or ions. Second, implement it as a parallel code. This is used to 
compute the properties, such as the excitation energies, 
hyperfine structure constants, oscillator strengths and, more importantly,
the life time of $^3P^o_0$ clock state, associated with the $^1S_0 - ^3\!P^o_0$ 
clock transition in Al$^+$. And, third, examine in the detail the 
role and contributions of triple excitations, Breit interaction and 
QED corrections in these properties. 

The remaining part of the paper is divided into five sections. In Sec. II, 
we discuss the FSRCC method for two-valence atomic systems. The properties 
calculation using two-valence FSRCC and contributing diagrams are discussed 
in Sec. III. The results obtained from our calculations are discussed and 
analyzed in Sec. IV. In Sec. V, we discuss the theoretical uncertainty of
our results. Unless stated otherwise, all results and equations 
presented in this paper are in atomic units ( $\hbar=m_e=e=1/4\pi\epsilon_0=1$).


\section{Two-valence FSRCC}

The wavefunction of a two-valence atom or ion, $|\Psi_{vw}\rangle$, is the 
solution of the eigenvalue equation
\begin{equation}
  H^{\rm DCB}|\Psi_{vw} \rangle = E_{vw} |\Psi_{vw} \rangle,
  \label{hdc_2v}
\end{equation}
where $E_{vw}$ is the exact energy. The Hamiltonian $H^{\rm DCB}$ 
is the Dirac-Coulomb-Breit no-virtual-pair Hamiltonian 
\begin{eqnarray}
   H^{\rm DCB} & = & \sum_{i=1}^N \left [c\bm{\alpha}_i \cdot
        \mathbf{p}_i + (\beta_i -1)c^2 - V_{N}(r_i) \right ]
                       \nonumber \\
    & & + \sum_{i<j}\left [ \frac{1}{r_{ij}}  + g^{\rm B}(r_{ij}) \right ],
  \label{ham_dcb}
\end{eqnarray}
where $\bm{\alpha}$ and $\beta$ are the Dirac matrices, and the last two 
terms, $1/r_{ij} $ and $g^{\rm B}(r_{ij})$, are the Coulomb and Breit 
interactions, respectively.
In FSRCC, $|\Psi_{vw} \rangle$ can be written as
\begin{equation}
|\Psi_{vw}\rangle = e^T \left[ 1 + S_1 + S_2 + 
	\frac{1}{2} \left({S_1}^2 + {S_2}^2 \right) + 
	R\right ]|\Phi_{vw}\rangle.
  \label{2v_exact}
\end{equation}
Here, $vw\ldots$ represent the valence orbitals and 
$|\Phi_{vw}\rangle, = a^\dagger_wa^\dagger_v |\Phi_0\rangle,$
is the Dirac-Fock reference state for the two-valence atom or ion. And,
$T$, $S$ and $R$ are coupled-cluster (CC) operators for the closed-shell, 
one-valence and two-valence sectors of the Hilbert space of the total 
electrons.

For a two-valence system with $N$-electrons, $T$, $S$ and $R$ operators 
in principle can have all possible excitations of the electrons, and therefore 
can be expressed as
\begin{equation}
	T= \sum_{i=1}^{N-2} T_i, \;\; S= \sum_{i=1}^{N-1} S_i, \;\; {\rm and} 
       \;\; R= \sum^N_{i=1} R_i.
\end{equation}
However, among all the excitations, the single and double subsume 
most of the electron correlation effects. And, therefore, we can approximate 
$T=T_1 + T_2$, $S=S_1 + S_2$ and $R=R_2$, this is referred to as the
coupled-cluster with singles and doubles (CCSD) approximation. The dominant
contributions from the triple excitations are, however, also included in the
present work using the perturbative triples approach, discussed later in
the paper. In the second quantized notation, these operators can be 
represented as
\begin{subequations}
\begin{eqnarray}
   T_1  = \sum_{ap}t_a^p a_p^{\dagger}a_a {\;\; \rm and \;\;} 
   T_2  = \frac{1}{2!}\sum_{abpq}t_{ab}^{pq} a_p^{\dagger}a_q^{\dagger}a_ba_a,
\end{eqnarray}
\begin{eqnarray}
   S_1 = \sum_{p}s_v^p a_p^{\dagger}a_v  {\;\; \rm and \;\;}
   S_2 = \sum_{apq}s_{va}^{pq} a_p^{\dagger}a_q^{\dagger}a_aa_v,
\end{eqnarray}
\begin{eqnarray}
  R_2 = \sum_{pq}r_{vw}^{pq} a_p^{\dagger}a_q^{\dagger}a_wa_v.
\end{eqnarray}
 \label{t1t2}
\end{subequations}
Here, the indices $ab\ldots$ and $pq\ldots$ represent the core and virtual 
orbitals, respectively. And, $t_{\ldots}^{\ldots}$, $s_{\ldots}^{\ldots}$ 
and $r_{\ldots}^{\ldots}$ are the cluster amplitudes corresponding to $T$, 
$S$ and $R$ CC operators, respectively.

The closed-shell and one-valence CC operators are obtained by solving 
the set of coupled nonlinear equations discussed in our previous works
Refs. \cite{mani-09} and \cite{mani-10}, respectively. Moreover, the 
details related to the computational implementation of RCC method for 
closed-shell and one-valence systems is given in our work \cite{mani-17}, 
where we had reported the details of our RCC codes. The two-valence CC 
operator $R_2$ is the solution of the equation \cite{mani-11}
\begin{eqnarray}
   \langle\Phi^{pq}_{vw}|
    \bar H_{\rm N} +
   \{\contraction{\bar}{H}{_{\rm N}}{S}\bar H_{\rm N}S^{'}\} +
   \{\contraction{\bar}{H}{_{\rm N}}{S}\bar H_{\rm N}R_2\}
   |\Phi_{vw}\rangle = \nonumber \\
   E^{\rm att}_{vw}
   \langle\Phi^{pq}_{vw}|\Bigl[S^{'} + R_2 \Bigr]|\Phi_{vw}\rangle.
   \label{ccsd_2v}
\end{eqnarray}
Here, for compact notation we have use
$S'=S^{(1)}_1+S^{(1)}_2 + \frac{1}{2}( {S^{(1)}_1}^2 + {S^{(1)}_2}^2)$. 
$E^{\rm att}_{vw}$ is the two-electron attachment energy and it is the
difference between the correlated energy of $(n-2)-$electron (closed-shell)
sector and $n-$electron (two-valence) sector, $E_{vw} - E_0$. Alternatively, 
it can also be expressed as
\begin{equation}
  E^{\rm att}_{vw} = \epsilon_v + \epsilon_w + \Delta E^{\rm att}_{vw},
  \label{2v_eatt}
\end{equation}
where $\epsilon_v$ and $\epsilon_w$ are the Dirac-Fock energy of the valence
electrons in $|\phi_v\rangle$ and $|\phi_w\rangle$, respectively. And,
$\Delta E^{\rm att}_{vw},  = \Delta E^{\rm corr}_{vw} -
\Delta E^{\rm corr}_0$, is the difference of the correlation energies of
closed-shell and two-valence sectors.

\section{Properties Calculation using FSRCC}

\subsection{Hyperfine matrix elements}

In this section we describe the properties calculation using the two-valence 
FSRCC. For a detailed discussion we consider the matrix elements of the 
hyperfine interaction. The approach, however, is also applicable for calculation 
of properties associated with other one-body operators with appropriate
selection rules. The hyperfine interaction is the coupling between the 
nuclear electromagnetic moments and the electromagnetic fields of the 
electrons. And, the  hyperfine interaction Hamiltonian \cite{johnson-07} is
\begin{equation}
  H_{\rm hfs} = \sum_i\sum_{k, q}(-1)^q t^k_q(\hat {\bf r}_i) T^k_{-q},
  \label{hfs_ham}
\end{equation}
where $t^k_q(\mathbf{r})$ and $T^k_{q}$ are the irreducible tensor operators
of rank $k$ in the electronic and nuclear sectors, respectively. 

Using the two-valence RCC wave function from Eq. (\ref{2v_exact}), the 
hyperfine matrix element in the electronic sector is
\begin{eqnarray}
&&\langle\Psi_i|H^{\rm e}_{\rm hfs}|\Psi_j\rangle = \sum_{k l}{c^i_k}^*c^j_l
    \left[\langle\Phi_k|\widetilde H^{\rm e}_{\rm hfs}
  + \widetilde H^{\rm e}_{\rm hfs} \Big(S^{'}
               \right. \nonumber \\
  &&\left. + R_2 \Big )
    + \left(S^{'} + R_2 \right )^\dagger \widetilde H^{\rm e}_{\rm hfs}
    + \left(S^{'} + R_2 \right )^\dagger
               \right. \nonumber \\
  &&\left.\; \widetilde H^{\rm e}_{\rm hfs}\left(S^{'} + R_2 \right )
    |\Phi_l\rangle \right],
  \label{hfs_cc}
\end{eqnarray}
where, $H^{\rm e}_{\rm hfs}$ is the electronic component of the
hyperfine operator. And, for compact notation, we
represent the two-valence state $|\Psi_{vw}\rangle$ with $|\Psi_{i}\rangle$.
The constants $c^i_j$ are the mixing coefficients corresponding to the
configuration state function $|\Phi_j\rangle$ for the state $|\Psi_i\rangle$, 
and are obtained by diagonalizing the effective Hamiltonian 
matrix \cite{mani-11} within the chosen model space.
The dressed hyperfine Hamiltonian
$\widetilde H^{\rm e}_{\rm hfs}= e{^T}^\dagger H^{\rm e}_{\rm hfs} e^T,$
is a non terminating series of closed-shell CC operator $T$. In our previous
work \cite{mani-10} we proposed an iterative scheme to include a class of 
dominant diagrams to all orders of $T$ in $\widetilde H^{\rm e}_{\rm hfs}$. 
And, we also showed that the terms cubic in $T$ and higher contribute less than 
$0.1\%$ to the properties. So, in the present work, we truncate 
$\widetilde H^{\rm e}_{\rm hfs}$ to second-order in $T$ and include the terms
$ \widetilde H^{\rm e}_{\rm hfs} \approx H^{\rm e}_{\rm hfs}
 + H^{\rm e}_{\rm hfs}T + T^\dagger H^{\rm e}_{\rm hfs}
 + T^\dagger H_{\rm hfs}T$ in the properties calculations.

Next, to assess the contributions from different sectors we group the terms
in Eq. (\ref{hfs_cc}) as
\begin{eqnarray}
  \langle\Psi_i|H^{\rm e}_{\rm hfs}|\Psi_j\rangle &=&
  \langle\Psi_i|H^{\rm e}_{\rm hfs}|\Psi_j\rangle_{\rm DF}
  + \langle\Psi_i|H^{\rm e}_{\rm hfs}|\Psi_j\rangle_{\rm 1v} \nonumber \\
  &&+ \langle\Psi_i|H^{\rm e}_{\rm hfs}|\Psi_j\rangle_{\rm 2v}.
  \label{hfs_cc1}
\end{eqnarray}
Here, the first, second, and third terms denote the contributions from the 
Dirac-Fock, one-valence, and two-valence sectors, respectively. The CC terms
arising from each of the  sectors are discussed in more detail. 
%
%
\begin{figure}[h]
\begin{center}
  \includegraphics[width = 8.0 cm]{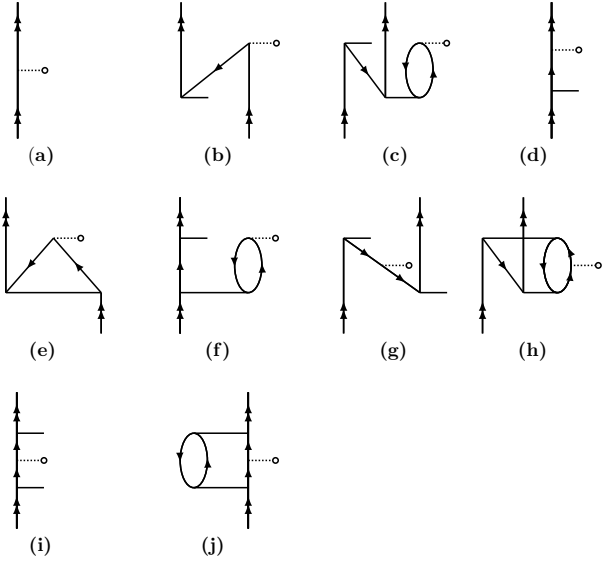}
	\caption{(a) The DF diagram. (b-j) Some contributing example diagrams 
	to Eq. (\ref{hfs_1v}). The diagrams are given in the same sequence as 
	the terms in Eq. (\ref{hfs_1v})}.
  \label{1vd}
\end{center}
\end{figure}


\subsubsection{Dirac-Fock contribution}
\begin{figure}
\begin{center}
  \includegraphics[width = 8.5cm]{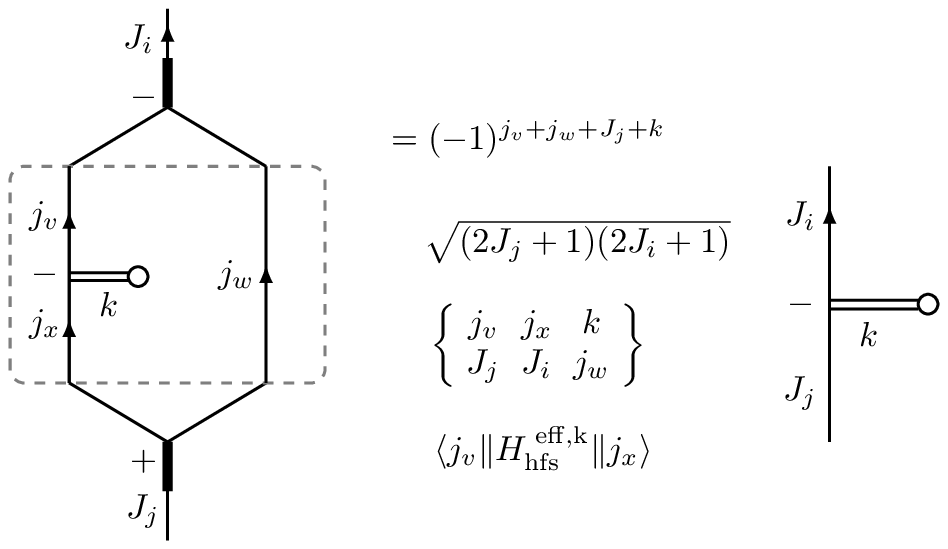}
  \caption{Angular factor arising from the coupling of one-body
           effective operator and a spectator valence line.
	   The free diagram on the right-hand side represents the
	   geometrical part in the Wigner-Eckart theorem.}
  \label{ang1v}
\end{center}
\end{figure}

The Dirac-Fock term is expected to have the dominant contribution among the 
three terms in Eq. (\ref{hfs_cc1}). It is the expectation of the bare 
hyperfine Hamiltonian operator 
\begin{equation}
  \langle\Psi_i|H^{\rm e}_{\rm hfs}|\Psi_j\rangle_{\rm DF} = \sum_{k l}{c^i_k}
   ^*c^j_l \langle\Phi_k| H^{\rm e}_{\rm hfs} |\Phi_l\rangle.
  \label{hfs_df}
\end{equation}
In terms of Goldstone diagrams, it has only one diagram and it is shown
in Fig. \ref{1vd}(a). Since $H^{\rm e}_{\rm hfs}$ is an one-body operator,
the contribution is the expectation of $H^{\rm e}_{\rm hfs}$
with respect to a valence orbital and then coupled with a spectator
valence orbital. The angular momentum diagram from the coupling is 
topologically equivalent to the one in Fig. \ref{ang1v} with the effective 
operator $H^{\rm eff, k}_{\rm hfs}$ replaced by $H^{k}_{\rm hfs}$. The 
labels $j_v$, $j_w$, $\ldots$, $(J_i,J_j)$ denote the angular momentum 
quantum numbers of uncoupled (coupled) states, and multipole $k$ represents 
the rank of the hyperfine operator.
%
%

\subsubsection{$\langle\Psi_i|H^{\rm e}_{\rm hfs}|\Psi_i\rangle_{\rm 1v}$
              contribution}

The contribution in this sector involves both the $T$ and $S$ operators. From 
Eq. (\ref{hfs_cc}) we can write
\begin{widetext}
\begin{eqnarray}
  \langle\Psi_i|H^{\rm e}_{\rm hfs}|\Psi_j\rangle_{\rm 1v}&= &\sum_{k l}{c^i_k}
   ^*c^j_l \left[\langle\Phi_k|\left( H^{\rm e}_{\rm hfs} T_1 + T^\dagger_1 H^
   {\rm e}_{\rm hfs} T_2 + \widetilde H^{\rm e}_{\rm hfs} S_1 
   + \widetilde H^{\rm e}_{\rm hfs} S_2 + {S_1}^\dagger 
   \widetilde H^{\rm e}_{\rm hfs} {S_2} \right ) + {\rm h.c.}                          
   + T^\dagger_1 H^{\rm e}_{\rm hfs} T_1 \right .
	                            \nonumber \\
	&&\left . + T^\dagger_2 H^{\rm e}_{\rm hfs} T_2 +   
	{S_1}^\dagger \widetilde H^{\rm e}_{\rm hfs} S_1  +
	{S_2}^\dagger \widetilde H^{\rm e}_{\rm hfs} S_2
	|\Phi_l\rangle \right] 
  \label{hfs_1v}
\end{eqnarray}
\end{widetext}
The above terms lead to 64 Goldstone diagrams and example diagrams
are shown Fig. \ref{1vd}(b-j). The leading order contribution is expected 
from $\widetilde H^{\rm e}_{\rm hfs} S$ and its hermitian conjugate 
$S^\dagger\widetilde H^{\rm e}_{\rm hfs}$. The example diagrams 
of $\tilde H^{\rm e}_{\rm hfs} S$ are shown in Fig. \ref{1vd}(d) and (e). The 
next leading order contribution is expected to be from the terms with two 
orders of $S$ operators, $S^\dagger\widetilde H^{\rm e}_{\rm hfs}S$. The 
example diagrams corresponding to this term are shown in 
Fig. \ref{1vd}(f), (i) and (j). To compute the contribution from
$\langle\Psi_i|H^{\rm e}_{\rm hfs}|\Psi_j\rangle_{\rm 1v}$, first we compute
the matrix elements with respect to uncoupled states and store them in
the form of an one-body effective operator. And then, like in the DF, this
effective operator is coupled with a spectator valence state.


\subsubsection{$\langle\Psi_i|H^{\rm e}_{\rm hfs}|\Psi_i\rangle_{\rm 2v}$
               contribution}

\begin{figure}
\begin{center}
  \includegraphics[width = 8.0cm]{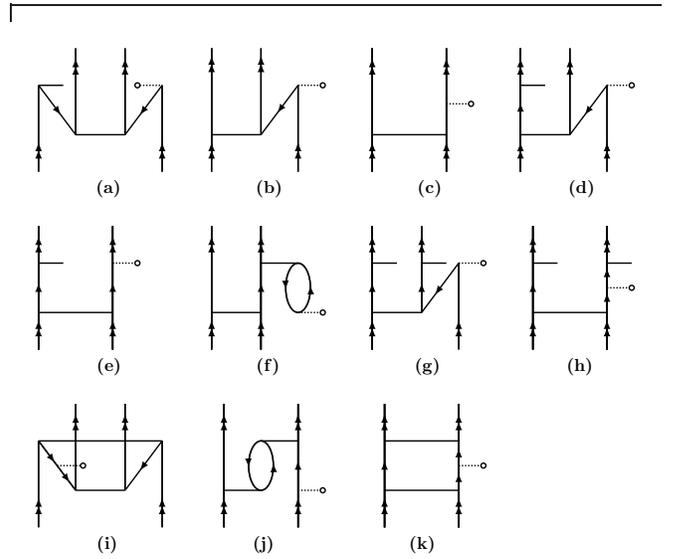}
  \caption{Some contributing example diagrams to Eq. (\ref{hfs_2v}).
	   For easy identification, diagrams are given in the same
	   sequence as the terms in Eq. (\ref{hfs_2v})}
  \label{2vd}
\end{center}
\end{figure}

\begin{figure}
\begin{center}
  \includegraphics[width = 8.5cm]{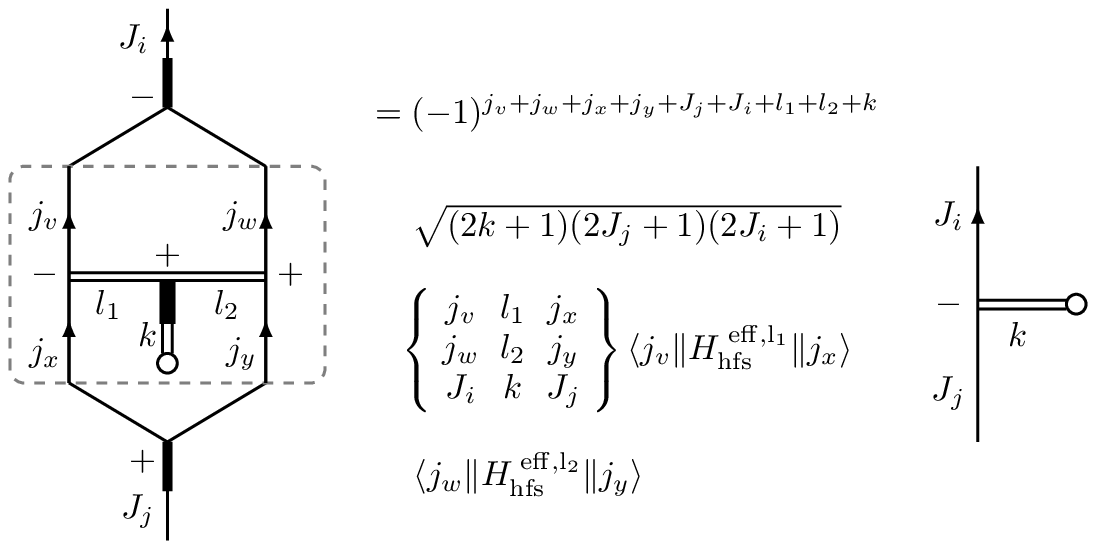}
  \caption{Angular factor arising from the coupling of two-body
           effective operator. Portion in the dashed rectangle
           is an effective operator which subsumes the contribution
           from Eq. (\ref{hfs_2v}) in terms of uncoupled states.}
  \label{ang2v}
\end{center}
\end{figure}

This term has contributions from all types of CC operators, $T$, $S$ and $R$,
\begin{widetext}
\begin{eqnarray}
  \langle\Psi_i|H^{\rm e}_{\rm hfs}|\Psi_j\rangle_{\rm 2v}& = &
    \sum_{k l}{c^i_k}^*c^j_l \left[\langle\Phi_k
    |\left(T^\dagger_1 H^{\rm e}_{\rm hfs} T_2 
    +\widetilde H^{\rm e}_{\rm hfs} S_2
    + \widetilde H^{\rm e}_{\rm hfs} R_2 + {S_1}^\dagger
    \widetilde H^{\rm e}_{\rm hfs} {S_2} + {(S_1 + S_2)}^\dagger 
    \widetilde H^{\rm e}_{\rm hfs} {R_2}
    + {S^2_1}^\dagger \widetilde H^{\rm e}_{\rm hfs} (S_2 + R_2)\right) \right .
                     \nonumber  \\
   &&\left .+ {\rm h.c.} + T^\dagger_2 H^{\rm e}_{\rm hfs} T_2 
    + {S_2}^\dagger \widetilde H^{\rm e}_{\rm hfs} S_2
    + {R_2}^\dagger \widetilde H^{\rm e}_{\rm hfs} R_2 |\Phi_l\rangle \right].
  \label{hfs_2v}
\end{eqnarray}
\end{widetext}
Here, we have neglected the terms with more than two-orders in $S_2$ 
as these will have negligible contribution. 
There are 68 diagrams which arise from this term. Like in the one-valence
sector, we give selected diagrams from this term in Fig. \ref{2vd} as example. 
The leading order contribution is expected to be 
$\widetilde H^{\rm e}_{\rm hfs}R_2$ and its hermitian conjugate
${R_2}^\dagger\widetilde H^{\rm e}_{\rm hfs}$. The corresponding example 
diagram from these terms is shown in Fig. \ref{2vd}(c). This is on account of 
two important reasons. First, these are the lowest order terms in $R_2$. 
And second, the magnitude of $R_2$ is larger than the $T$ and $S$. The next 
leading order contribution is expected to be $\widetilde H^{\rm e}_{\rm hfs}S_2$  
and its hermitian conjugate as these are one order in $S$. The corresponding
example diagram is shown in Fig. \ref{2vd}(b). Among the terms which second
or higher order in CC operators, the dominant contribution is expected from 
the term ${R_2}^\dagger\tilde H^{\rm e}_{\rm hfs}R_2$. Diagrammatically, an
example is shown in Fig. \ref{2vd}(k). The reason for this is attributed to 
the larger magnitudes of $R_2$ operators. The remaining terms are expected to 
have negligible contributions.  To compute the contribution from 
$\langle\Psi_i|H^{\rm e}_{\rm hfs}|\Psi_i\rangle_{\rm 2v}$, all the
terms in Eq. (\ref{hfs_2v}) are computed with respect to uncoupled states
first and then stored in the form of a two-body effective operator, as shown 
in Fig. \ref{ang2v}. And, as indicated in the figure, the angular momenta of 
the valence electrons are coupled.

%
%

\subsection{Contribution from perturbative $R_3$}

\begin{figure}
\begin{center}
\includegraphics[width = 8.4cm]{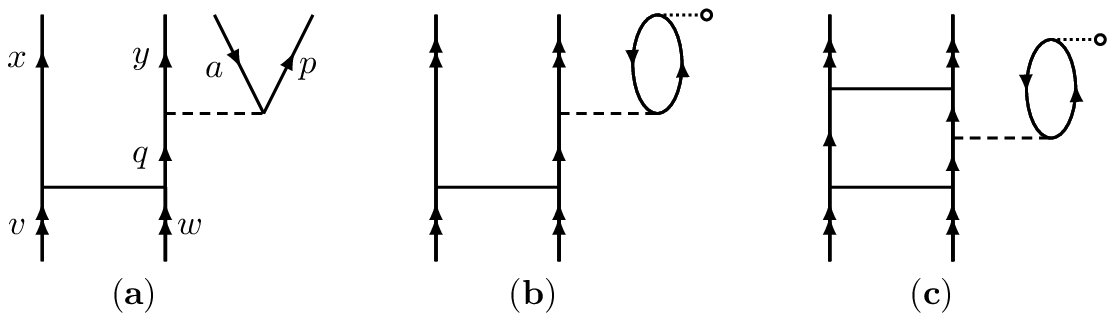}
	\caption{(a) The perturbative $R_3$ diagram.
	(b-c) The hyperfine matrix element diagrams from the terms
        $H^e_{\rm hfs}R_3$ and $R_2^\dagger H^e_{\rm hfs}R_3$.
        Dashed line represents the two-body residual interaction,
	$g_{ij}$, between the electrons.}
  \label{ptrbr3}
\end{center}
\end{figure}

To account for the electron correlation effects from triple excitations,
we consider the perturbative triples.  With this approach we can
incorporate the dominant contributions from triple excitations, however, with
far less computational cost than the full triples. For this, we choose the
triples which arise from the two-valence CC operator $R_2$, and the
term is $\contraction[0.5ex]{}{g}{}{R} gR_2$,
where $g_{ij} = \sum_{i<j}[\frac{1}{r_{ij}} + g^{\rm B}(r_{ij})]$, the two-body
residual interaction. This has the leading order contribution to triples, 
since the magnitude  of $R_2$ is larger than $T$ and $S$ for two-valence 
systems. The diagram corresponding to $\contraction[0.5ex]{}{g}{}{R} gR_2$ is
shown in Fig. \ref{ptrbr3} (a), and the algebraic expression is
\begin{equation}
	R_3 \approx \frac{1}{\Delta\epsilon^{xyp}_{vwa}}
	a^\dagger_x a^\dagger_y a^\dagger_p a_a a_w a_v
	\sum_{q}\langle yp|g|qa\rangle\langle xq|R_2|vw\rangle,
\end{equation}
where $\Delta\epsilon^{xyp}_{vwa} = \epsilon_{v}+\epsilon_{w}+\epsilon_{a}
-\epsilon_{x}-\epsilon_{y}-\epsilon_{p}$. The operator $R_3$
contract with other CC operators along with the hyperfine operator and
contribute to the properties through Eq. (\ref{hfs_cc}). In our previous work 
on the two-valence systems \cite{mani-11}, the dominant contribution to 
the properties involves the cluster operator $R_2$. So, in the present work, to 
account for the contribution from $R_3$ we include the terms 
$H^e_{\rm hfs}R_3$, $R_3^\dagger H^e_{\rm hfs}$,
$R_2^\dagger H^e_{\rm hfs}R_3$ and $R_3^\dagger H^e_{\rm hfs}R_2$.
There are 3 diagrams from each of these terms which contribute to
the two-valence properties. And, as example, one diagram each from the 
terms $H^e_{\rm hfs}R_3$ and and $R_2^\dagger H^e_{\rm hfs}R_3$ are shown
in Fig. \ref{ptrbr3} (b) and (c), respectively.

\subsection{Hyperfine induced E1 transition}

The hyperfine eigenstate $|\Gamma FM_F\rangle$ is obtained by coupling the 
electronic state $|\Psi_{vw}\rangle$ with the eigenstate of the nuclear 
spin $I$. Considering the hyperfine interaction $H_{\rm hfs}$ as a 
perturbation and using the first-order time-independent perturbation theory
\begin{eqnarray}
	|\Gamma F M_F \rangle =  \sum_n && \left [
  \frac{\langle \gamma_n J_n \gamma_I I|H_{\rm hfs} |
                \gamma_0 J_0 \gamma_I I \rangle} 
        {E_{J_0}-E_{J_n}} \right ] \nonumber \\
   && \times |\gamma_n J_n \gamma_I I \rangle. 
  \label{psi_hfs}
\end{eqnarray}
The term within the parenthesis represents the hyperfine mixing of 
unperturbed state $|\gamma_0 J_0 \gamma_I I \rangle$ state with an excited 
state $|\gamma_n J_n \gamma_I I \rangle$. The parameters $\Gamma$ and 
$\gamma_i$ are additional quantum numbers to identify the states 
uniquely, and $E_J$ is the exact energy. The transition amplitude between 
two hyperfine states $|\Gamma_i F_i M_{F_i} \rangle$ and 
$|\Gamma_j F_j M_{F_j} \rangle$ is 
\begin{equation}
	E1_{\rm HFS} = \langle \Gamma_iF_i||\mathbf{D}||\Gamma_j F_j\rangle,
	\label{e1hfs}
\end{equation}
where $\mathbf{D}$ is the electric dipole operator. Using the expression
for $|\Gamma F M_F \rangle$ from Eq. (\ref{psi_hfs}) in the above equation, 
we obtain the expression for the $E1_{\rm HFS}$ induced 
$^1S_0-^{3}P^o_0$ transition amplitude as
\begin{eqnarray}
        E1_{\rm HFS} & = & c(I,J,F,\mu_I) \left [ \frac{\langle {^1}S_0||d||{^3}P^o_1\rangle
                          \langle {^3}P^o_1||t^1||{^3}P^o_0\rangle}
                         {\Delta E_{{^3}P^{o}_1}} \right.  \nonumber \\
              && \left.  +       \frac{\langle {^1}S_0||d||{^1}P^o_1 \rangle
                          \langle {^1}P^o_1||t^1||{^3}P^o_0 \rangle}
                         {\Delta E_{{^1}P^{o}_1}} \right ],
\label{e1hfs2}
\end{eqnarray}
where $c(I,J,F,\mu_I)$ is the angular factor associated with hyperfine
wavefunction in Eq. (\ref{psi_hfs}). And, $\Delta E_{{^3}P^{o}_1}$ and 
$\Delta E_{{^1}P^{o}_1}$ are the energy differences 
$E_{{^3}P^{o}_0}-E_{{^3}P^{o}_1}$ and $E_{{^3}P^{o}_0}-E_{{^1}P^{o}_1}$.

\section{Results and Discussions}

\subsection{Convergence of basis}

To obtain accurate results it is crucial to use a basis set which provides 
a good description of the single-electron wave functions and 
energies. And, to incorporate the effects of finite charge distribution of 
the nucleus we use a two-parameter finite size Fermi density distribution.
In this work, we use the Gaussian-type orbitals (GTOs) \cite{mohanty-91} as 
the single-electron basis. The orbital as well as the self-consistent-field 
energies are optimized to match the GRASP2K \cite{jonsson-13} data. We 
achieve excellent match and details of the comparison is reported in our 
recent work \cite{ravi-20}. The orbital basis used in the present work also 
incorporates the effects of Breit interaction, vacuum polarization and the 
self-energy corrections. For Breit interaction,
we employ the expression given in Ref. \cite{grant-80} and incorporate it 
in the orbital generation as well as the FSRCC calculations. The effect 
of vacuum polarization to single-electron orbitals is considered using 
the Uehling potential \cite{uehling-35} modified for the finite size
nucleus \cite{johnson-01}. The self-energy corrections to the 
orbitals are incorporated through the model Lamb-shift operator introduced by 
Shabaev {\em et al.} \cite{shabaev-13}, and are calculated using the code 
QEDMOD \cite{shabaev-15}. 

\begin{figure}
\begin{center}
\includegraphics[width = 8.2cm, angle=-90]{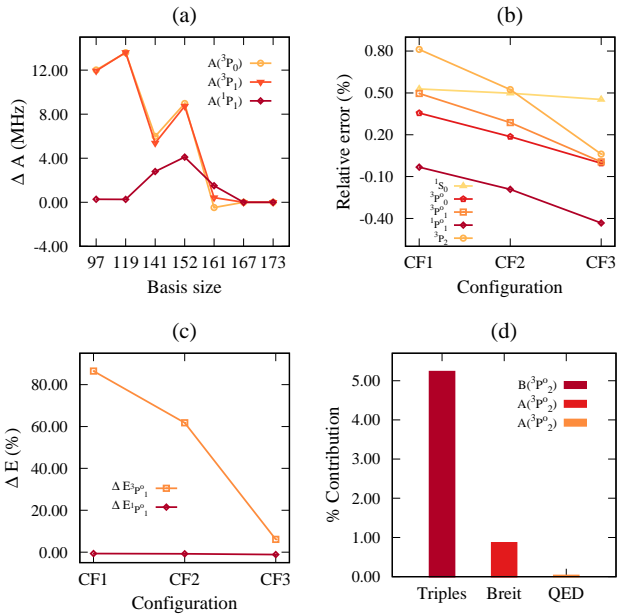}
	\caption{The convergence trend of HFS constants as function of basis 
                 size (panel (a)), relative errors in the excitation energy 
                 and the energy separation as function of configurations 
                 (panel (b) and (c), respectively), and maximum percentage 
                 contributions from the perturbative triples, 
		 Breit interaction and QED corrections to HFS constants.}
  \label{deltaE}
\end{center}
\end{figure}

Mathematically, the GTO basis are incomplete \cite{grant-06} and hence, it 
is essential to check the convergence of results with basis size. For this, 
we start with a moderate basis of 86 orbitals ($14s$, $14p$, $9d$, $5f$, 
$4g$, $4h$) and add orbitals in each symmetry until the change in the 
properties is $\leqslant 10^{-3}$ in respective units of the properties. For 
illustrative purposes the convergence trend of the magnetic dipole hyperfine 
structure constant is shown in Fig. \ref{deltaE}(a). It is observed that 
the change is less than $10^{-3}$ MHz when the basis is augmented from 
167 to 173. So, to optimize the compute time, we consider the basis set 
with 167 ($23s$, $23p$, $15d$, $12f$, $11g$, $11h$) orbitals as optimal, 
and use it in the properties computations.

\subsection{ Excitation energies}

\begin{table}
	\caption{Energy (in cm$^{-1}$) of the ground state $3s^2\ ^1S_0$ 
	and the excitation energies of low-lying excited states using the 
	configurations $3s^2+3s3p+3p^2+3s3d+3s4s$ in the model space. 
	Listed energies also incorporate the contributions from the 
	Breit interaction and QED corrections, and are obtained using 
	the converged basis of 167 orbitals.}
  \label{tab-ee}
  \begin{ruledtabular}
  \begin{tabular}{lccc}
	  States &  FSRCC & Other cal. & Exp. \cite{nist} \\
	  \hline
 $3s^2\ ^1S_0$  & 379582& $381210^{\rm j}$, $381331^{\rm d}$, $381287^{\rm k}$ & 381308 \\
                &          & $382024^{\rm c}$                                     &           \\
 $3s3p\ ^3P^o_0$& 37395& $37392^{\rm j}$, $37374^{\rm k}$, $37396^{\rm d}$  &  37393  \\
                &          & $37191^{\rm c}$                                    &           \\
 $3s3p\ ^3P^o_1$& 37452& $36705^{\rm a}$, $35000^{\rm b}$, $37454^{\rm j}$ &  37454  \\
                &          & $36292^{\rm l}$, $37516^{\rm m}$, $37457^{\rm d}$ &            \\
                &          & $37818^{\rm n}$, $37253^{\rm p}$, $37251^{\rm c}$ &            \\
 $3s3p\ ^3P^o_2$& 37555& $37579^{\rm j}$, $37572^{\rm d}$, $37374^{\rm c}$ &  37578  \\
 $3s3p\ ^1P^o_1$& 60111& $60723^{\rm a}$, $63000^{\rm b}$, $59855^{\rm j}$  &  59852  \\
                &          & $59849^{\rm k}$,$59427^{\rm l}$, $60198^{\rm m}$  &            \\
                &          & $59768^{\rm d}$, $59140^{\rm n}$, $60104^{\rm p}$ &            \\
                &          & $54410^{\rm c}$                                    &           \\
 $3p^2\ ^1D_2$&   85578& $85450^{\rm j}$,  $85462^{\rm d}$, $85678^{\rm c}$ &  85481  \\
 $3s4s\ ^3S_1$&   91043& $91256^{\rm j}$,  $91289^{\rm d}$, $91262^{\rm k}$  &  91274  \\
 $3p^2\ ^3P_0$&   93379& $94049^{\rm j}$,  $94092^{\rm d}$, $93672^{\rm c}$  &  94085  \\
 $3p^2\ ^3P_1$&   93380& $94112^{\rm j}$,  $94151^{\rm d}$, $93735^{\rm c}$  &  94147  \\
 $3p^2\ ^3P_2$&   93409& $94234^{\rm j}$,  $94265^{\rm d}$, $93857^{\rm c}$  &  94269  \\
 $3s4s\ ^1S_0$&   95156& $95336^{\rm j}$,  $95354^{\rm d}$  &  95350.60  \\
 $3s3d\ ^3D_1$&   95248& $95420^{\rm j}$,  $95527^{\rm d}$, $95532^{\rm k}$   &  95551  \\
              &            & $95695^{\rm c}$                                    &           \\
 $3s3d\ ^3D_2$&   95252& $95419^{\rm j}$,  $95527^{\rm d}$, $95697^{\rm c}$  &  95550  \\
 $3s3d\ ^3D_3$&   95253& $95418^{\rm j}$,  $95524^{\rm d}$, $95690^{\rm c}$  &  95549  \\
 $3s3d\ ^1D_2$&  110382& $106270^{\rm c}$ & 110090  \\
 $3p^2\ ^1S_0$&  111598&  & 111637  \\
  \end{tabular}
  \end{ruledtabular}
\begin{tabbing}
  $^{\rm j}$Ref.\cite{konovalova-15}[CI+AO], \
  $^{\rm k}$Ref.\cite{mitroy-09}[CICP], \     
  $^{\rm d}$Ref.\cite{safronova-11}[CI+AO],  \\ 
  $^{\rm a}$Ref.\cite{stanek-96}[CIDF+CP], \  
  $^{\rm b}$Ref.\cite{das-90}[MCDF], \
  $^{\rm l}$Ref.\cite{chou-93a}[MCRRPA],     \\ 
  $^{\rm m}$Ref.\cite{zou-2000}[MCDHF], \      
  $^{\rm n}$Ref.\cite{safronova-2000}[RMBPT], \
  $^{\rm p}$Ref.\cite{jonsson-97}[MCDF], \\    
  $^{\rm c}$Ref.\cite{johnson-97}[RMBPT]
\end{tabbing}
\end{table}

In Table \ref{tab-ee}, we list the low-lying energies of Al$^+$ from our 
results along with other theory and experimental data for comparison. From 
the table it is evident that our results are in good agreement with 
experimental as well as previous theoretical results. 
The largest and smallest relative errors in our calculation are 0.9\% 
and 0.004\%, in the case of $3p^2{\;^3}P_{2}$ and $3s3p{\;^3}P^{o}_0$ states, 
respectively. It is to be noted that the states with low energy configurations 
$3s^2$ and $3s3p$, which are key to clock transition, 
are very close to the experiment. Among the previous theoretical results, 
those from the CI+AO calculations by Konovalova and 
collaborators \cite{konovalova-15}, and Safronova and 
collaborators \cite{safronova-11} are in better agreement with the 
experimental data. 
The maximum relative error is $\approx$ 0.14\% in each of these calculations, 
in the case of $3s3d{\;^3}D_{2}$ and $3s3p{\;^1}P^o_{1}$ states, respectively.
The reason for the marginal difference between these calculations and 
ours can be attributed to the different treatment of {\em core-core} and 
{\em core-valence} correlations. In Refs. \cite{konovalova-15} and 
\cite{safronova-11}, a linearized CCSD is used in the calculation. 
However, in the present work, we include the nonlinear terms in the CCSD.
Hence, our work consider the electron correlation effects better than the
previous works. This naturally translates to improved over all uncertainty.
The other set of reliable results, in terms of proximity to the experimental 
data, are based on the CICP method obtained by Mitroy and 
collaborators \cite{mitroy-09}. The remaining theoretical results are either 
based on many-body perturbation theory or multi-configuration Hartree-Fock 
and these have larger deviations from the experiment. Considering the 
contributions from the Breit and QED corrections, we observe the largest 
combined contribution of $\approx0.01$\% of the total value in the case 
of $^3P^o_0$. The magnitude is consistent with the previous 
calculation \cite{konovalova-15}.

To discern the electron correlation effects, the energies are computed with 
three different model spaces. We start with the configurations 
$3s^2+3s3p$ (CF1) in the model space and then add $3s4s$ and $3p^2+3s3d$ in 
the two subsequent computations CF2 and CF3, respectively. We could not 
separate the contribution 
from $3p^2$ and $3s3d$ as the inclusion of any one of these leads to 
divergence due to the {\em intruder} states. To elaborate, when only $3p^2$ 
is included in the model space, $3s3d{\;^3}D_{1,2,3}$ states, having
energies within the range of the model space, are the intruder states. This 
leads to divergence due to small energy denominator. Similarly, when only 
$3s3d$ is included, $3p^2{\;^3}P_{0,1,2}$ states are intruder states.
The trends of the results from the three model spaces are shown in 
Fig. \ref{deltaE}(b) and (c). The plots in the figures show that 
the inclusion of $3p^2$, $3s3d$ and $3s4s$ in the model space improves
the energies of the $3s^2{\;^1}S_{0}$ and $3s3p\; ^3P^{o}_{0,1,2}$ states, and 
the energy difference
$\Delta E_{{^3}P^{o}_1}$, $= E_{{^3}P^{o}_0}-E_{{^3}P^{o}_1}$. 
Obtaining correct value for the energy difference $\Delta E_{{^3}P^{o}_1}$ 
is the key to obtain accurate life time of the ${^3}P^{o}_0$ state. 
The improvement can be attributed to the inclusion of {\em valence-valence} 
correlation effects more accurately by diagonalizing the effective 
Hamiltonian in a larger model space. As a result, $\Delta E_{{^3}P^{o}_1}$ 
increases from 8.24 ${\rm cm}^{-1}$ to 57.76 ${\rm cm}^{-1}$, which is in 
good agreement with the experimental result of 60.88 ${\rm cm}^{-1}$. This 
improves the life time of the ${^3}P^{o}_0$ state  by about 96\%. 
We, however, observe an opposite trend for the excitation energy 
of the ${^1}P^{o}_{1}$ state and hence, the value of the 
energy difference $\Delta E_{{^1}P^{o}_1}$ as well.  But these have 
negligible effect on the life time of ${^3}P^{o}_0$ as 
$\Delta E_{{^1}P^{o}_1}$ is very large, $\approx$ 22457 ${\rm cm}^{-1}$. 
It must be emphasized that the two-valence coupled-cluster calculations
with larger model space is challenging. And, our present work demonstrates
the possibility of doing this with FSRCC without ambiguity by augmenting the 
model space systematically.

\subsection{Hyperfine and dipole reduced matrix elements 
            and structure constants}

\begin{table*}
\centering
\caption{Magnetic dipole and electric quadrupole hyperfine structure constants
	(in MHz) for ${^3}P^o_1$, ${^3}P^o_2$ and ${^1}P^o_1$ states. The
	values of the nuclear magnetic dipole moment $\mu_I = 3.6415069(7) 
	\mu_N$ and electric quadrupole moment $Q = 0.1466(10){\rm b}$ are used 
	in the calculation.}
\label{hfs-con}
\begin{ruledtabular}
\begin{tabular}{lrrrrrr}
Methods & \multicolumn{6}{c}{Hyperfine Structure Constants} \\
	\hline
	& \multicolumn{3}{c}{$A$} & \multicolumn{3}{c}{$B$} \\
          \cline{2-4}  \cline{5-7}
        & ${^3}P^o_1$ & ${^3}P^o_2$ & ${^1}P^o_1$ & ${^3}P^o_1$ & ${^3}P^o_2$ & ${^1}P^o_1$ \\
	\hline
CCSD        &  $1385.409$    &   $1188.024$   &   $292.588$     &    $-16.173$      &  $25.549$      &  $27.876$      \\
CCSD(T)     &  $   8.316$    &   $  -3.865$   &   $ -8.661$     &    $ -0.473$      &  $-1.276$      &  $ 0.432$      \\
Breit       &  $  -4.240$    &   $ -10.194$   &   $  2.153  $   &    $ -1.633[-4]$  &  $ 1.246[-3]$  &  $-2.745[-4]$  \\
Vacuum pol. &  $   0.306$    &   $   0.306$   &   $ -4.118[-4]$ &    $  4.069[-5]$  &  $-3.715[-5]$  &  $-1.956[-4]$  \\
Self-energy &  $   0.023$    &   $   0.023$   &   $  3.790[-4]$ &    $  2.561[-6]$  &  $ 4.323[-5]$  &  $ 3.966[-5]$  \\
Total       &  $1389.814$    &   $1174.294$   &   $286.0800$    &    $-16.646$      &  $24.274$      &  $28.308$      \\
Other cal.  &  $1348^{\rm a}$ & $1149^{\rm a}$ & & $-15.62^{\rm a}$ & $31.42^{\rm a}$ & \\
\end{tabular}
\end{ruledtabular}
\begin{tabbing}
  $^{\rm a}$Ref.\cite{itano-2007}[MCDHF], \
\end{tabbing}
\end{table*}

The magnetic dipole and electric quadrupole hyperfine constants, $A$ and $B$, 
respectively, obtained from our study are listed in the Table \ref{hfs-con}. 
In addition, the off-diagonal reduced matrix elements required to evaluate 
the $E1_{\rm HFS}$ amplitude are presented in the Table \ref{e1-amp}. For 
quantitative assessment the contributions from the Breit interactions, QED 
corrections and dominant triples are also listed in the table. For all the 
states CCSD is the dominant contribution. And, the DF term has the leading 
order contribution among the different sectors in the CCSD, it accounts for 
more than 90\% of the total value. More importantly, within $2v$, 
$H^{\rm e}_{\rm hfs}R_2 + {\rm h.c.}$ has the largest contribution. This can 
be attributed to the larger magnitude of the $R_2$ operator. As discernible 
from the Fig. \ref{deltaE}(d), the contribution from the perturbative 
triples $R_3$ is also crucial. For example, it has $\approx$ 3 and 5\% of 
the total value of $A ({^1}P^{o}_1)$ and $B ({^3}P^{o}_2)$, respectively. 
This implies that the triples must be included in the FSRCC calculations to 
obtain accurate results for hyperfine structure and related properties of 
Al$^+$. From the Breit interaction, $A ({^3}P^{o}_2)$ has the largest 
contribution $\approx$ 0.9\%. Considering the level of the accuracy needed for 
clock properties, it is a significant contribution and can not be neglected. 
The contribution from the QED corrections is $\approx$ 0.02\% and negligible
compared to the other terms.

\begin{table}
\caption{Magnetic dipole hyperfine and E1 transition reduced matrix 
	elements, $t_{30}=\langle {^3}P^o_1||t^1||{^3}P^o_0\rangle$
	and $t_{10}=\langle {^1}P^o_1||t^1||{^3}P^o_0 \rangle$, 
      $d_{03}=\langle {^1}S_0||d||{^3}P^o_1\rangle$ and  
	$d_{01}=\langle {^1}S_0||d||{^1}P^o_1 \rangle$, 
	in atomic units.}
  \label{e1-amp}
\begin{ruledtabular}
\begin{tabular}{lrrrr}
	Methods &  $d_{03}$ & $d_{01}$ & $t_{30}$ & $t_{10}$  \\
                        \hline
CCSD        & $-1.425[-2]$ & $ 2.841$     & $ -0.095   $ & $ 0.079$ \\
CCSD(T)     & $-9.384[-4]$ & $-9.159[-4]$ & $  1.608[-3]$ & $-5.667[-4]$ \\
Breit       & $ 6.303[-5]$ & $-2.069[-5]$ & $  4.882[-5]$ & $-3.165[-4]$ \\
Vacuum pol. & $-9.737[-7]$ & $1.181[-5] $ & $ -4.198[-5]$ & $ 3.035[-5]$ \\
Self-energy & $ 2.332[-7]$ & $5.608[-7]$  & $ -2.993[-6]$ & $ 2.849[-6]$ \\
Total       & $-1.513[-2]$ & $2.840$      & $ -0.094   $ & $ 0.078$ \\
Other cal.  & &  &  $-0.120^{\rm a}$ & $0.096^{\rm a}$ \\
            & &  &  $-0.119^{\rm b}$ &  \\
\end{tabular}
\end{ruledtabular}
\begin{tabbing}
 $^{\rm a}$Ref.\cite{kang-09}[MCDHF], \
 $^{\rm b}$Ref.\cite{beloy-17}[CI+MBPT]
\end{tabbing}
\end{table}

To the best our knowledge, there are no experimental data for comparison. 
However,  there is one theoretical result each for $A$ and $B$ of 
${^3}P^{o}_1$ and ${^3}P^{o}_2$ using the MCDF method by Itano and 
collaborators \cite{itano-2007}. Our results of 1389.81, 1174.29 and 16.65 
for $A({^3}P^{o}_1)$, $A({^3}P^{o}_2)$ and $B({^3}P^{o}_1)$, respectively 
are $\approx 3.1$, $2.2$ and $5.6$\%, larger than the values given in the 
Ref. \cite{itano-2007}. The reason for this difference can be attributed to 
the better accounting of the electron correlations in FSRCC theory as it 
includes the residual Coulomb interaction to all orders.  We observe an 
opposite trend for $B({^3}P^{o}_2)$. Our DF result 29.44 is close to the 
MCDF result 31.42 \cite{itano-2007}, but our total value of 24.23 is 
45.7\% lower. This is due to the large cancellation from the $2v$ sector. 
For the off-diagonal reduced matrix elements, there are two previous results 
for comparison. The magnitudes of the reduced matrix elements $t_{30}$ and 
$t_{10}$ from our calculation are smaller than the MCDHF \cite{kang-09} and 
CI+MBPT \cite{beloy-17} results. The reason for this can be attributed to the 
difference in the treatment of the electron correlations in FSRCC and 
these calculations.
For the results in Ref. \cite{kang-09}, there could be two sources of 
uncertainty in the matrix elements. First, the active space of CSFs is
limited to single-electron basis with $n=7$ and $l=5$ only, where $n$ and $l$ 
are principal and orbital quantum numbers, respectively. And second, the 
core polarization effect is considered only from the $2s$ and $2p$ electrons. 
In the present work, however, we include a large active space with orbitals 
up to $n= 25$ and $l=6$, and CSFs arising from all core-to-valence, 
valence-to-virtuals and core-to-virtuals single and double electron 
replacements. In addition, we also include the contribution from triple 
excitations perturbatively. Coming to the results in Ref. \cite{beloy-17}, 
there is an important difference in terms of accounting the {\em core-core} 
and {\em core-valence} correlations. In the present work, these are 
considered up to all orders of residual Coulomb interaction. In
Ref. \cite{beloy-17}, however, considers these up to third-order only.
It is to be also mentioned that the uncertainty in the reduced matrix 
elements in Ref. \cite{beloy-17} is 3\%.


We use the E1 transition reduced matrix elements in Table \ref{e1-amp}
to compute the oscillator strengths, which are listed in the 
Table \ref{tab-os}. 
Like hyperfine structure constants, the dominant contribution is from the CCSD. 
It contributes more than 94\% of the total value. The contributions from the 
perturbative triples and Breit interactions are significant. The maximum 
contributions from these are $\approx$ 6.2 and 0.4\%, respectively. 
Like the case of hyperfine, QED correction has a negligible contribution.
For the oscillator strength of the $^1S_0-^3P^o_1$ transition there is one 
experimental result and it is based on the time-resolved 
technique \cite{johnson-86}. Our theoretical result, 
$2.60\times 10^{-5}$ a.u., for this transition has the same order of 
magnitude as the experimental data $(1.068 \pm 0.074)\times 10^{-5}$ a.u., 
but $\approx$ 128\% larger. The other theoretical results, although based on 
MCDF or related methods, show wide variation. The results range from 
$0.36\times 10^{-5}$ \cite{das-90} to 
$3.78 \times 10^{-5}$ a.u. \cite{stanek-96}. For the ${^1}S_0-{^1}P^o_1$ 
transition there are three experimental results based on the beam-foil 
technique \cite{Kernahan_1979,baudinet-79,berry_70}. Despite the same
experimental technique, there is a large variation in the results. In 
addition, the uncertainties associated with the results are large, these are 
in the range $\approx$ 4.8\% \cite{baudinet-79} to 
15.8\% \cite{Kernahan_1979, berry_70}. Our theoretical result 1.47 lies 
within the range of the experimental values. One observation is that the 
previous theoretical results are similar in values. The reason for this 
could be the similar treatment of the electron correlations as all are 
based on MCDF and its variations. And, have similar
shortcomings in the inclusion of electron-correlation effects. This highlights
the importance of cross checking with other methods like we have done with
a better method.
\begin{table}
\caption{Oscillator strengths of the allowed transitions compared
    with other calculations and experiments.
    Here, $[x]$ represents $10^x$.}
  \label{tab-os}
\begin{ruledtabular}
\begin{tabular}{lcc}
 Method & $^1S_0 - ^3P^o_1$ & $^1S_0 -  ^1P^o_1$    \\
                                                  \hline
CCSD(T)+    & $2.604[-5]$ & $1.473$     \\
Bre.+QED  &  &     \\
Other cal. & $3.560[-6]^{\rm a}$, $8.875[-6]^{\rm b}$, & $1.740^{\rm a}$, $1.765^{\rm e}$, $1.831^{\rm b}$,  \\
           & $3.776[-5]^{\rm c}$, $1.017[-5]^{\rm d}$  & $1.850^{\rm f}$, $1.746^{\rm g}$, $1.751^{\rm h}$ , \\
           &                                           & $1.76^{\rm c}$ , $1.775^{\rm d}$   \\
Expt.  &   $(1.068\pm0.074)[-5]^{\rm i}$  & $1.74\pm 0.27^{\rm j}$, $1.9\pm 0.3^{\rm l}$ \\
        &                   & $1.26 \pm 0.06^{\rm k}$   \\
\end{tabular}
\end{ruledtabular}
\begin{tabbing}
  $^{\rm a}$Ref.\cite{das-90}[MCDF],\
  $^{\rm b}$Ref.\cite{chou-93a}[MCRRPA],\
  $^{\rm c}$Ref.\cite{stanek-96}[MCDF+CP], \\
  $^{\rm d}$Ref.\cite{kang-09}[MCDHF],\
  $^{\rm e}$Ref.\cite{mitroy-09}[CICP],\
  $^{\rm f}$Ref.\cite{shorer-77}[RRPA], \\
  $^{\rm g}$Ref.\cite{zou-2000}[MCDHF], \
  $^{\rm h}$Ref.\cite{jonsson-97}[MCDF],\
  $^{\rm i}$Ref.\cite{johnson-86}[Exp.], \\
  $^{\rm j}$Ref.\cite{Kernahan_1979}[Exp.],\
  $^{\rm k}$Ref.\cite{baudinet-79}[Exp.],\
  $^{\rm l}$Ref.\cite{berry_70}[Exp.]
\end{tabbing}
\end{table}

\subsection{$E1_{\rm HFS}$}

\begin{table}
\centering
	\caption{Wavelength ($\lambda$) (in nm), $E1_{\rm HFS}$ 
	amplitude (in a.u.) of $^1S_0-^3P^o_0$ transition and the life 
	time ($\tau$) (in sec.) of $^3P^o_0$ metastable state. 
	Here, $[x]$ represents $10^x$. }
\label{tab-e1hfs}
\begin{ruledtabular}
\begin{tabular}{lccc}
	Methods  & $\lambda$& $E1_{\rm HFS}$ & $\tau$    \\          
	                           \hline  
	CCSD     & 267.44  & $5.153[-5]$ & $21.33$ \\
	CCSD(T)  &         & $5.316[-5]$ & $20.04$  \\
	CCSD(T)+Bre.+QED  &  & $5.295[-5]$   & $20.20\pm0.91$   \\
	Other cal.  & &                      & $23.11^{\rm a}$, 
	                                       $20.33^{\rm b}$  \\  
	Exp.    & 267.43  &               & $20.6\pm1.4^{\rm c}$ \\ 
\end{tabular}
\end{ruledtabular}
\begin{tabbing} 
$^{\rm a}$ Ref.\cite{kang-09}[MCDF], \
$^{\rm b}$ Ref.\cite{brage-98}[MCDF], \
$^{\rm c}$ Ref.\cite{rosenband-07}[Exp.] \ 
\end{tabbing}
\end{table}
Using the electric dipole and hyperfine reduced matrix elements from
Table \ref{e1-amp} and the energy differences $\Delta E_{^{3}P^o_1/^{1}P^o_1}$
from Table \ref{tab-ee} in Eq. (\ref{e1hfs2}), we calculate the
$E1_{\rm HFS}$ amplitude of $^1S_0-^{3}P^o_0$ transition and the life time 
of the $^{3}P^o_0$ clock state. The results from the present and previous
works are listed in the Table \ref{tab-e1hfs}. 
The experimental value of the life time is 
$20.6\pm1.4$ s from Ref. \cite{rosenband-07}. This is in very good agreement 
with our theoretical value $20.20\pm0.91$ s identified as CCSD(T)+Breit+QED in 
the table. Here, one point is to be noted, the error associated with the
experimental value $\approx$ 6.8\% is not negligible. As discernible from the 
table, the contribution from the perturbative triples to the life time 
is $\approx -6.4$\% of the total value, and is essential to 
improve the comparison with the experimental result. The combined contribution 
from the Breit interaction and QED corrections is $\approx$ 0.8\% of the 
total value. Considering the current uncertainties of optical atomic clocks, 
this cannot be neglected to obtain theoretical results with commensurate
uncertainties. Two previous theoretical works, using the MCDF method, have 
reported the life time of the $^{3}P^o_0$ state \cite{brage-98, kang-09}. 
Among the two, the recent work Ref. \cite{kang-09} treats the electron 
correlation more accurately by considering {\em single} and {\em double} 
electron replacements and larger active space for CSFs. However, the result of
23.11 s in Ref. \cite{kang-09} has a larger deviation ($\approx$ 12\%) from the 
experimental data. This indicates inherent shortcomings or inconsistencies of 
accounting the electron correlation properly in the MCDF method. This is 
resolved in the present work. In the FSRCC method such inconsistencies do not 
arise. 
As mentioned earlier, we use a converged basis as active space
in which all possible {\em single} and {\em double} electron replacements
are included to all orders.


\section{Theoretical Uncertainty} 

The theoretical uncertainty in the lifetime of the $^3P^o_0$ state
depends on the uncertainties in the HFS reduced matrix elements $t_{30}$ 
and $t_{10}$, the dipole reduced matrix elements $d_{03}$ and $d_{01}$, and 
the energy denominators $\Delta E_{{^3}P^{o}_1}$ and $\Delta E_{{^1}P^{o}_1}$.
For the reduced matrix elements, we have identified four sources which 
contribute to the theoretical uncertainty. First source the truncation of the 
basis set. As shown in Fig. \ref{deltaE}(a) for HFS constant, the change in 
the HFS and electric dipole matrix elements is of the order of $10^{-3}$ or
less on augmenting the converged basis. Since the change is very small, we 
can neglect this uncertainty. Second source is the truncation 
of the dressed Hamiltonian $\tilde{H}^{\rm e}_{\rm hfs}$ to second order in 
$T^{(0)}$. In our previous work on hyperfine structure 
constants \cite{mani-10}, using an iterative scheme, we have shown that the 
contribution from third and higher order terms is less than 0.1\%.
So, we take 0.1\% as an upper bound from this source of uncertainty.
Third source is the partial inclusion of triple excitations in the 
properties calculation. Since we consider the leading order terms of triple 
excitation in the perturbative triples, the contribution from remaining 
terms will be small. Based on our analysis in present and previous 
works \cite{ravi-20, chattopadhyay-15} we estimate the upper bound from 
this source as 0.72\%.
Fourth source of uncertainty is associated with the frequency-dependent
Breit interaction which is not included in the present work. However,
in our previous work \cite{chattopadhyay-14} using a series of computations
with GRASP2K which implements this interaction we estimated an upper bound
on this uncertainty to be $0.13$\% in Ra. Although Al$^+$ is much lighter
atom and expected to have much smaller contribution from frequency-dependent
Breit interaction, we take $0.13$\% as an upper bound from this source.
There could be other sources of theoretical uncertainty, such as the higher
order coupled perturbation of vacuum polarization and self-energy terms,
quadruply excited cluster operators, etc. But, these, in general, have
much lower contributions to the properties and their cumulative
theoretical uncertainty could be below 0.1\%. 
The theoretical uncertainty associated with energy denominators 
$\Delta E_{{^3}P^{o}_1}$ and $\Delta E_{{^1}P^{o}_1}$ are calculated from 
the relative errors in the excitation energies of $^3P^{o}_0$, $^3P^{o}_1$ 
and $^1P^{o}_1$ states. These are $\approx$ 0.01\% and 0.43\%, respectively. 
 
The other theoretical uncertainty which will contribute to the life time is 
the QED corrections at the level of $E1_{\rm HFS}$ calculation. 
To estimate this uncertainty, we refer to Refs. \cite{shabaev-05a, shabaev-05b}. 
In these works Shabaev and collaborators have implemented and computed the one-loop 
QED corrections to the parity-nonconserving transition amplitudes. Considering
that the magnetic dipole HFS like the matrix element of the parity violating
interaction Hamiltonian, the associated theoretical uncertainty would be
similar. Thus, based on these works we consider 0.3\% as the upper bound from 
this source of uncertainty. It is, however, to be noted that the actual
uncertainty would be smaller as Al$^+$ is a much lighter system compared
to the Cs and Fr atoms studied in Refs. \cite{shabaev-05a, shabaev-05b}.
So, by combining the upper bounds of all the contributions, the
theoretical uncertainty associated with the value of the life time 
of ${^3}P^o_0$ state is 4.5\%.

\section{Conclusions}

In conclusion, we have developed an all particle Fock-space relativistic 
coupled-cluster based method to calculate the properties of two-valence 
atomic systems. To account for the relativistic effects and QED corrections we 
use the Dirac-Coulomb-Breit Hamiltonian with the corrections from 
the Uehling potential and the self-energy. The effects of the triple 
excitations are incorporated using the perturbative triples. Using the method 
we have calculated the properties such as the excitation energies, hyperfine 
structure constants and reduced matrix elements, oscillator strengths, and the 
life time of the $^3P^o_0$ metastable state in Al$^+$, which is 
an important parameter for the $^1S_0 - ^3P^o_0$ clock transition. 
Our results of the excitation energies and oscillator strengths agrees well 
with the experimental data. Most importantly, our theoretical estimate of 
the life time of the $^3P^o_0$ state, $20.20 \pm 0.91$ s, is in excellent 
agreement with the experimental value, $20.60 \pm 1.4$ s, from Rosenband {\it et al.} 
\cite{rosenband-07}. From our studies we conclude that the contributions from 
the triple excitations and Breit+QED corrections are essential to obtain 
accurate clock properties in Al$^+$. Based on error analysis, the upper 
bound on the theoretical uncertainty in the calculated life time of 
$^3P^o_0$ is 4.5\%. The level of uncertainty in our results indicates that the 
FSRCC method we have developed has the potential to predict the structure and 
properties of two-valence atoms and ions with accuracies commensurate with 
the experiments.

\begin{acknowledgments}
We would like to thank B P Das for useful suggestions on the manuscript.
One of the authors, BKM, acknowledges the funding support from the 
SERB (ECR/2016/001454). The results presented in the paper are based on the 
computations using the High Performance Computing cluster, Padum, at the Indian 
Institute of Technology Delhi, New Delhi.
\end{acknowledgments}

\appendix
\section{Convergence table of the properties with basis size}

In Table \ref{energy_conv}, we provide the trend of the convergence
of excitation energies, magnetic dipole and electric quadrupole
hyperfine constants, and electric dipole transition amplitudes as
a function of basis size. As it is evident from the table, all the
properties converge to the order of $10^{-3}$ or less in the
respective units of the properties.

\begin{table*}
  \caption{Convergence of excitation energy, hyperfine structure constants
           and electric dipole transition amplitudes as function of basis size.}
  \label{energy_conv}
  \begin{ruledtabular}
  \begin{tabular}{lccccccccc}
          States/Property & \multicolumn{8}{c}{Basis size}  \\
  \hline
  & BS1 & BS2  & BS3 & BS4 & BS5 & BS6 & BS7 & BS8\\
  \hline
  & &&&{Exc. ene.} &&&& \\
  \cline{5-5}
 $3s3p\ ^3P^o_0$ & 36880.92 &  36887.02 &  36893.38 & 37050.55 & 37388.70 & 37391.43 &  37391.43 & 37391.43  \\
 $3s3p\ ^3P^o_1$ & 36941.51 &  36947.51 &  36953.75 & 37109.51 & 37448.96 & 37449.19 &  37449.19 & 37449.19  \\
 $3s3p\ ^3P^o_2$ & 37050.77 &  37056.56 &  37062.55 & 37215.38 & 37557.29 & 37552.51 &  37552.52 & 37552.52  \\
 $3s3p\ ^1P^o_1$ & 60174.87 &  60177.54 &  60181.35 & 60186.27 & 60205.24 & 60109.38 &  60109.38 & 60109.38  \\
 $3p^2\ ^1D_2$   & 84935.91 &  84943.85 &  84951.80 & 85142.41 & 85605.12 & 85574.35 &  85574.35 & 85574.35  \\
 $3s4s\ ^3S_1$   & 90488.94 &  90494.85 &  90500.30 & 90670.08 & 91019.53 & 91041.40 &  91041.40 & 91041.40  \\
 $3p^2\ ^3P_0$   & 92789.16 &  92801.94 &  92816.68 & 92977.64 & 93367.32 & 93374.60 &  93374.59 & 93374.59  \\
 $3p^2\ ^3P_1$   & 92798.39 &  92810.12 &  92823.73 & 92986.81 & 93380.10 & 93375.69 &  93375.69 & 93375.69  \\
 $3p^2\ ^3P_2$   & 92882.82 &  92894.07 &  92907.51 & 93053.81 & 93411.63 & 93405.18 &  93405.18 & 93405.18  \\
 $3s4s\ ^1S_0$   & 94524.80 &  94531.58 &  94537.68 & 94717.64 & 95132.51 & 95154.58 &  95154.59 & 95154.58  \\
 $3s3d\ ^3D_1$   & 94950.56 &  94948.62 &  94943.20 & 95034.90 & 95321.19 & 95246.39 &  95246.39 & 95246.39  \\
 $3s3d\ ^3D_2$   & 94954.45 &  94952.51 &  94947.56 & 95039.24 & 95325.07 & 95250.27 &  95250.27 & 95250.27  \\
 $3s3d\ ^3D_3$   & 94957.27 &  94955.31 &  94950.02 & 95041.37 & 95326.47 & 95251.68 &  95251.68 & 95251.68  \\
 $3s3d\ ^1D_2$   &109929.38 & 109931.55 & 109931.47 &110085.40 &110457.19 &110379.43 & 110379.43 & 110379.43\\
 $3p^2\ ^1S_0$   &111441.00 & 111448.22 & 111453.87 &111532.23 &111733.69 &111593.71 & 111593.71 & 111593.71\\ \\
          & &&&{HFS con.} &&&& \\
  \cline{5-5}
$A ({^3}P^o_1)$ & $1345.337$ & $1357.349$ & $1370.930$ & $1376.918$ & $1385.889$ & $1385.409$ & $1385.410$  &  $1385.410$  \\
$A ({^3}P^o_2)$ & $1147.931$ & $1159.867$ & $1173.480$ & $1178.863$ & $1187.596$ & $1188.024$ & $1188.025$  &  $1188.025$  \\
$A ({^1}P^o_1)$ & $ 283.646$ & $ 283.920$ & $ 284.187$ & $ 286.985$ & $ 291.088$ & $ 292.588$ & $ 292.588$  &  $ 292.588$  \\
$B ({^3}P^o_1)$ & $ -15.969$ & $ -15.978$ & $ -15.980$ & $ -16.059$ & $ -16.165$ & $ -16.173$ & $ -16.173$  &  $ -16.173$  \\
$B ({^3}P^o_2)$ & $  25.026$ & $  25.041$ & $  25.045$ & $  25.231$ & $  25.503$ & $  25.549$ & $  25.549$  &  $  25.549$  \\
$B ({^1}P^o_1)$ & $  27.340$ & $  27.355$ & $  27.358$ & $  27.576$ & $  27.825$ & $  27.876$ & $  27.876$  &  $  27.876$  \\ \\
 & &&&{E1 amp.} &&&& \\
 \cline{5-5}
${^1}S_0 - {^3}P^o_1$&$-1.843[-2]$&$-1.832[-2]$&$-1.820[-2]$ & $-1.718[-2]$&$-1.531[-2]$&$-1.425[-2]$ & $-1.425[-2]$ & $-1.425[-2]$ \\
          ${^1}S_0 - {^1}P^o_1$ & $2.894$ & $2.893$ & $2.893$ &  $2.875$ & $2.845$  & $2.841$  & $2.841$  & $2.841$ \\
  \end{tabular}
  \end{ruledtabular}
\footnotetext[1]{BS1 - 86 (14s, 14p,  9d,  5f,  4g,  4h)}
\footnotetext[2]{BS2 - 97 (15s, 15p, 10d,  6f,  5g,  5h)}
\footnotetext[3]{BS3 - 119 (17s, 17p, 12d,  8f,  7g,  7h)}
\footnotetext[4]{BS4 - 141 (19s, 19p, 14d, 10f,  9g,  9h)}
\footnotetext[5]{BS5 - 152 (20s, 20p, 15d, 11f, 10g, 10h)}
\footnotetext[6]{BS6 - 161 (21s, 21p, 15d, 12f, 11g, 11h)}
\footnotetext[7]{BS7 - 167 (23s, 23p, 15d, 12f, 11g, 11h)}
\footnotetext[8]{BS8 - 173 (25s, 25p, 15d, 12f, 11g, 11h)}
\end{table*}

\bibliography{al_2v}

\end{document}